\documentstyle[12pt,epsf,psfig]{article}
\def\be{\begin{equation}}
\def\ee{\end{equation}}
\def\bear{\begin{eqnarray}}
\def\eear{\end{eqnarray}}
\def\bea*{\begin{eqnarray*}}
\def\eea*{\end{eqnarray*}}
\def\bdm{\begin{displaymath}}
\def\edm{\end{displaymath}}

\def\G{\Gamma}
\def\d{\delta}

\def\l{\lambda}

\def\6{\partial}               
\def\7{\llap{/}}               

\def\nn{\nonumber}
\def\<{\langle}
\def\>{\rangle}

\makeatletter
\newbox\slashbox \setbox\slashbox=\hbox{\large$/$}
\def\pslash#1{\setbox\@tempboxa=\hbox{$#1$}
  \@tempdima=0.5\wd\slashbox \advance\@tempdima 0.5\wd\@tempboxa
  \copy\slashbox \kern-\@tempdima \box\@tempboxa}
\def\FMSlash{\protect\pslash}

\begin{document}

\title{Quantum Chaos in Physical Systems: from Super Conductors to Quarks}

\author{Elmar Bittner, Harald Markum, Rainer Pullirsch\\
\small{Institut f\"ur Kernphysik, Technische Universit\"at Wien,
A-1040 Vienna, Austria}}
\date{}\maketitle

\begin{abstract}
This article is the written version of a talk delivered at the Bexbach
Colloquium of Science 2000 and starts with an introduction into quantum chaos
and its relationship to classical chaos. The Bohigas-Giannoni-Schmit
conjecture is formulated and evaluated within random-matrix theory.
Several examples of physical systems exhibiting quantum chaos ranging
from nuclear to solid state physics are presented. The presentation concludes
with recent research work on quantum chromodynamics and the 
quark-gluon plasma. In the case of a chemical potential the eigenvalue 
spectrum becomes complex and one has to deal with non-Hermitian 
random-matrix theory.
\end{abstract}

\section{Classical Chaos}

Systems of classical mechanics can be divided into two classes: 
integrable and non-integrable systems. If there exists a
constant of motion $F(p,q)$ for the Hamiltonian $H(p,q)$ this
is connected with a symmetry
\bear
[H,F] = \frac{\partial H}{\partial p} \frac{\partial F}{\partial q} - 
\frac{\partial H}{\partial q} \frac{\partial F}{\partial p} \equiv 0 \ .
\eear
For a system with $n$ degrees of freedom with $n$ constants of motion
$F_i, \  i=1,...,n, $ and $[F_i,F_j] = 0$,
a canonical transformation to action-angle variables, $(p,q) \to (J,\theta)$,
can be performed
\bear
\frac{dJ_i}{dt}=-\frac{\partial H}{\partial \theta}= 0 \ , \mbox{\phantom{WWW}}
\frac{d \theta_i}{dt} = \frac{\partial H}{\partial J_i} = \omega_i \ ,
\eear
with frequencies $\omega_i$. The new equations of motions can be integrated
leading to
\bear
 \begin{array}{r@{\quad = \quad}l}
J_i & constant \\ \theta_i & \omega_i t + \varphi_i \ . \end{array}
\eear
Such a system is called integrable and each trajectory lies on an $n$-dimensional
torus. If this procedure is not possible, we have to deal with a non-integrable
systems. When one adds a non-integrable Hamiltonian $H_1$ to an integrable one $H_0$
\bear
H(J,\theta) = H_0(J) + \varepsilon H_1(J,\theta) \ ,
\eear
one can show that the system cannot be integrated by perturbation theory for 
rational frequency ratios
\bear
\frac{\omega_i}{\omega_j} = \frac{r}{s} \mbox{\phantom{WWW}} {\mbox{r,s}} \
\epsilon \ Z\! \ .
\eear    
The famous KAM theorem due to Kolmogorov, Arnold, and Moser states that
tori with $\frac{\omega_i}{\omega_j}$ sufficiently irrational are stable 
under small perturbations.
Non-integrable systems possess trajectories filling the phase space 
ergodically and are intrinsically related to chaotic motion. With the
KAM-theorem we get insight into the behavior of a system which is driven
by some (order) parameter from regularity to chaos~\cite{schuster}.

\section{Quantum Chaos}

It is a fascinating question in which manner classical chaos is reflected 
in quantum systems. Quantum mechanics is associated with the time dependent
Schr\"odinger equation
\begin{equation}
\hat{H} \psi = i \hbar \frac{\partial}{\partial t} \psi \: .
\end{equation}
In the stationary case
\begin{equation}
\psi(x,t) = \phi(x)\exp{(-i \omega t)}  
\end{equation}
one deals with the time-independent Schr\"odinger equation
\begin{equation}
- \frac{\hbar^2}{2m} \triangle \phi(x) + V(x) \phi (x) = E \phi (x) \: .
\end{equation}
This is a linear differential equation and its solution  behaves regularly in time.
Due to the Heisenberg uncertainty principle
\begin{equation}
\triangle p \triangle x \geq \frac{\hbar}{2} 
\end{equation}
the concept of trajectories is not adequate. Therefore, the question has be posed:
Are there differences in the eigenvalue spectra of classically integrable
and non-integrable systems?

\begin{figure}[ht] 
 \begin{center}
   \hspace{0.5cm}\psfig{figure=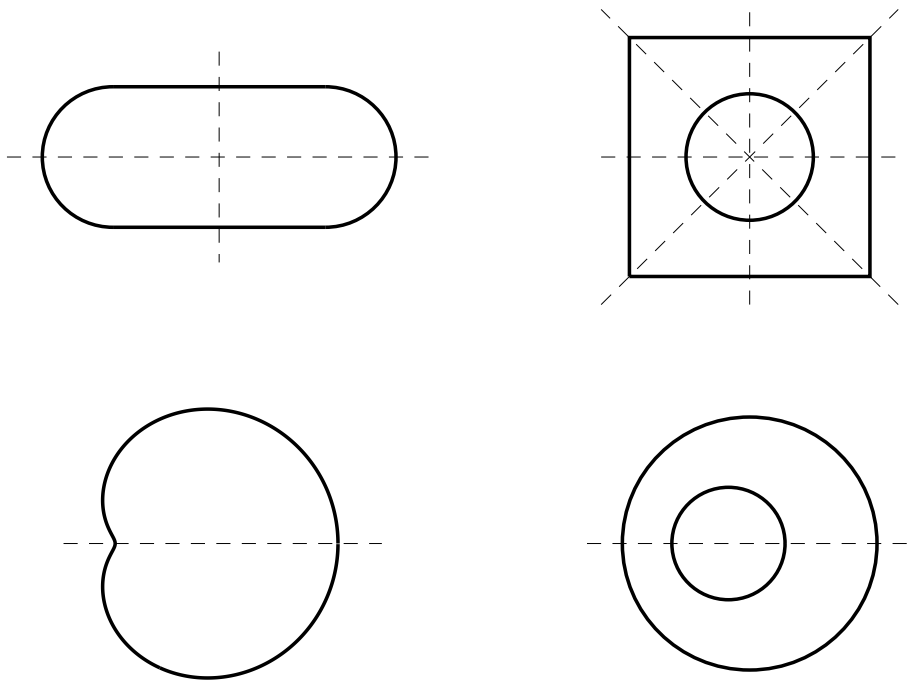,width=14cm,angle=-90}
   \end{center}
   \vspace{-1cm}
   \caption{Chaotic billiards: Bunimovich stadium (top left),
            Sinai billiard (top right), Pascalian snail (bottom
            left) and annular billiard (bottom right). Taken from
            Ref.~\cite{Guhr}.}
   \label{fig1}
 \end{figure}

\begin{figure}[t]
 \begin{center}
   \hspace{1cm}\psfig{figure=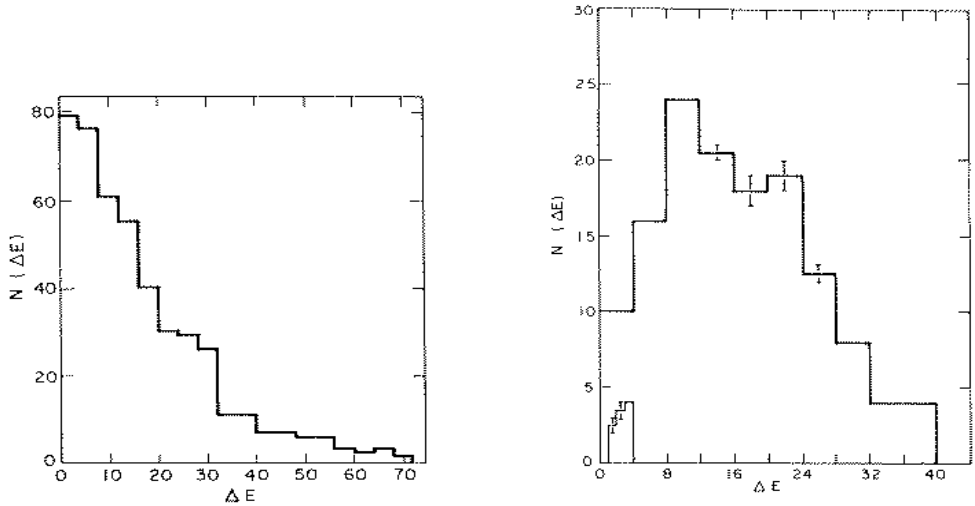,width=14cm,height=8cm}
   \end{center}
   \caption{Nearest-neighbor spacing distributions of eigenvalues
            for a circle (left) and the Bunimovich stadium (right).
            Taken from Ref.~\cite{McDoKauf}.}
   \label{fig2}
 \end{figure}

Billiards became a preferred playground to study both the classical
and quantum case. Some of the most important ones are depicted in
Fig.~\ref{fig1}. It was the arrival of computers with increasing
power in the late seventies when diagonalization of matrices with
reasonable size became possible. The behavior of the distribution
of the spacings between neighboring eigenvalues turned out to be a
decisive signature. In 1979 McDonald and Kaufman performed a
comparison between the spectra from a classically regular and a
classically chaotic system~\cite{McDoKauf}. As seen in Fig.~\ref{fig2}
they observed a qualitatively different behavior between the nearest-neighbor
spacing distribution of the circle and the stadium. In
the first case the spacings are clearly concentrated around zero
while they show repelling character in the second case.
There were several authors contributing to this discussion and we mention
the papers by Casati, Valz-Gris, and Guarneri~\cite{Casa}, by 
Berry~\cite{Berr}, by Robnik~\cite{Robn}
and by Seligman, Verbaarschot, and Zirnbauer~\cite{SeVeZi}.

\begin{figure}[h]
 \begin{center}
   \hspace{1cm}\psfig{figure=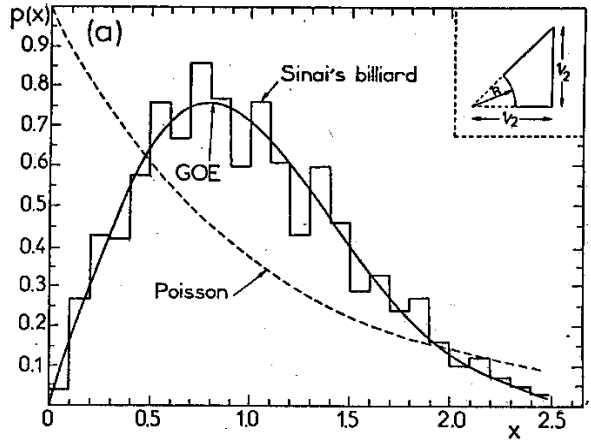,width=8cm,height=6cm}
   \end{center}
   \caption{Nearest-neighbor spacing distributions of eigenvalues for
            the Sinai billiard with the Wigner surmise compared to the
            Poisson distribution. The histogram comprises about 1000
            consecutive eigenvalues. Taken from Ref.~\cite{Bohi84}.}
   \label{fig3}
 \end{figure}

Very accurate results were obtained for the classically chaotic Sinai
billiard by Bohigas, Giannoni, and Schmit (see Fig.~\ref{fig3}) which
led them to the important conclusion~\cite{Bohi84}:
Spectra of time-reversal invariant systems whose classical analogues are
K systems show the same fluctuation properties as predicted by the Gaussian
orthogonal ensemble (GOE) of random-matrix theory (RMT).  K systems are
most strongly mixing classical systems with a positive Kolmogorov entropy.
The conjecture turned out valid also for less chaotic (ergodic) systems
without time-reversal invariance leading to the Gaussian unitary ensemble
(GUE).

\section{Random Matrix Theory}
\label{sec3}

In lack of analytical or numerical methods to obtain the spectra of complicated
Hamiltonians, Wigner and Dyson analyzed ensembles of random matrices and
were able to derive mathematical expressions.
A Gaussian random matrix ensemble consists of square matrices with their
matrix elements drawn from a Gaussian distribution
\begin{equation}
p(x) = \frac{1}{\sqrt{2\pi}\sigma} \; \exp\left(-\frac{x^2}{2\sigma^2}\right) \ .
\end{equation}
One distinguishes between three different types depending on space-time
symmetry classified by the Dyson parameter $\beta_D = 1, 2, 4$~\cite{Guhr}.
The Gaussian orthogonal ensemble (GOE, $\beta_D = 1$) holds for
time-reversal invariance and rotational symmetry of the Hamiltonian
\bear
H_{mn} = H_{nm} = H^{\ast}_{nm} \ .
\eear
When time-reversal invariance is violated and
\bear
H_{mn}=[H^{\dagger}]_{mn} \ ,
\eear
one obtains the Gaussian unitary ensemble (GUE, $\beta_D = 2$).
The Gaussian symplectic ensemble (GSE, $\beta_D = 4$) is in 
correspondence with time-reversal invariance but broken rotational
symmetry of the Hamiltonian
\bear
H^{(0)}_{nm}\mbox{1\hspace*{-2mm}I}_2 - i \sum\limits_{\gamma=1}^3 
H^{(\gamma)}_{nm} \sigma_{\gamma} \ ,
\eear
with $H^{(0)}$ real and symmetric and $H^{(\gamma)}$ real and antisymmetric.

The functional form of the distribution $P(s)$ of the neighbor
spacings $s$ between consecutive eigenvalues for the Gaussian
ensembles can be approximated by 
\bear
P_{\beta_D}(s) = a_{\beta_D} \, s^{\beta_D} \exp\left(-b_{\beta_D} \,
s^2 \,\right) \ ,
\eear
which is known as the Wigner surmise and reads for example in the case
$\beta_D =2$ (GUE)
\bear
P(s) = \frac{32}{\pi^2} \, s^2 \exp\left(-\frac{4}{\pi} \, s^2 \right) \ .
\eear
If the eigenvalues of a system are completely uncorrelated one ends up
with a Poisson distribution for their neighbor spacings
\bear
P(s) = \exp{(-s)} \ .
\eear
An interpolating function between the Poisson and the Wigner distribution
is given by the Brody distribution~\cite{Brod} reading for the GOE case
\bear
\label{brodydist}
P(s,\omega) = \alpha \, (\omega +1)  s^\omega 
\exp\left(- \alpha \, s^{\omega + 1}\right) \ ,
\quad \quad \alpha = \Gamma^{\omega+1} \,
\left(\frac{\omega + 2}{\omega + 1}\right) \ ,
\quad \quad 0 \leq \omega \leq 1 \ .
\eear

\section{Examples of Physical Systems}

So far the mathematical relationship between classically chaotic systems and
their eigenvalues after quantization. Remarkably, the Wigner distribution could
be observed in a number of systems by physical experiments and computer simulations
evading the whole quantum world~\cite{Guhr}.

\subsection{Atomic Nuclei}

An impressive manifestation of the Wigner surmise came from nuclear
physics~\cite{nuclear}.
\begin{figure}[hp]
 \begin{center}
   \hspace{1cm}\psfig{figure=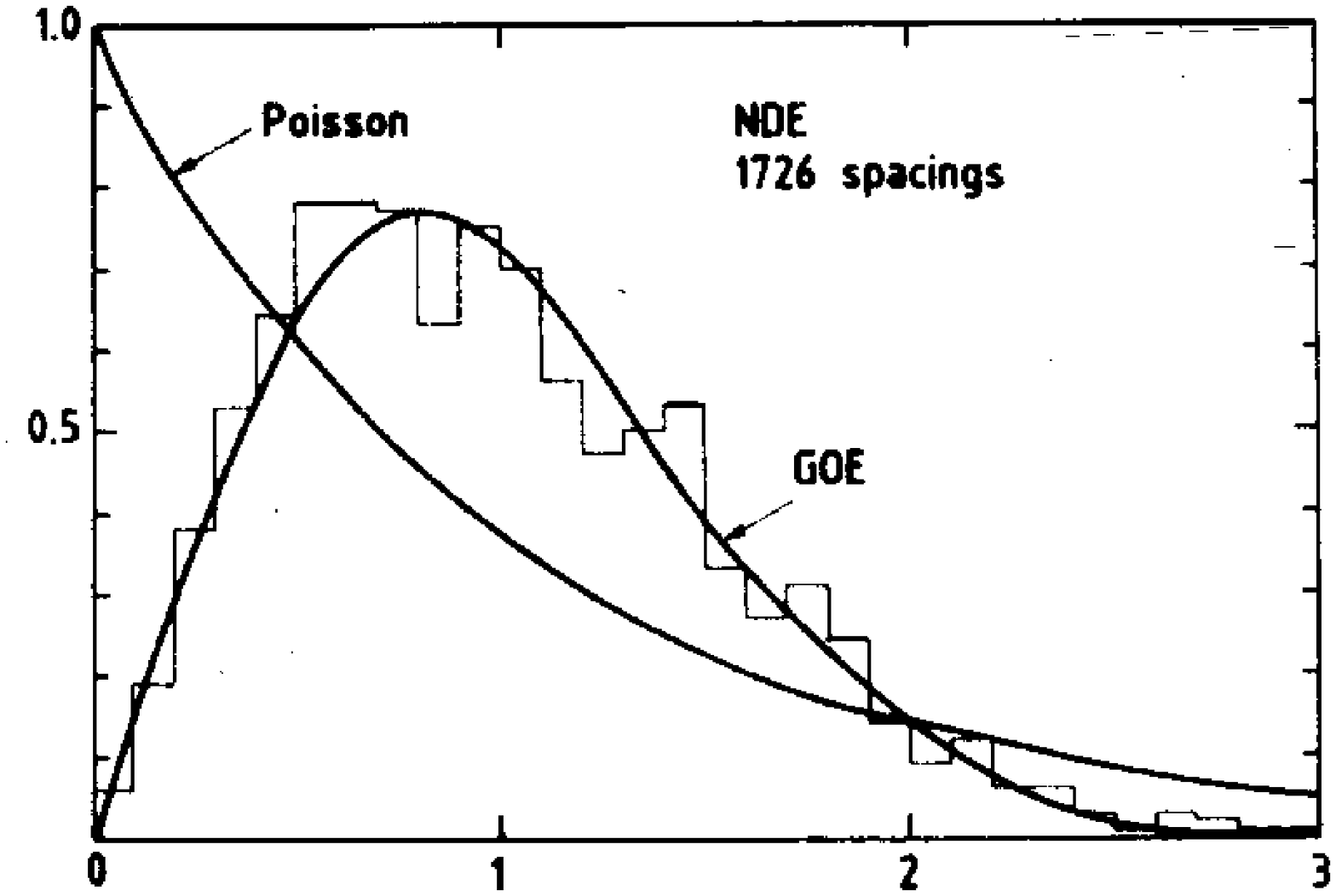,width=8.5cm,height=6cm}
   \end{center}
   \vspace{-4mm}
   \caption{Nearest-neighbor spacing distribution for a ``nuclear data ensemble''
            (NDE) compared to the RMT prediction labeled GOE and the Poisson
            distribution. Taken from Ref.~\cite{nuclear}.}
   \label{fig4}
 \begin{center}
   \hspace{1cm}\psfig{figure=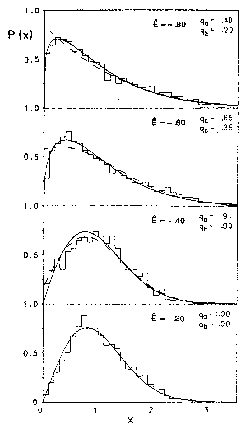,width=10cm,height=10.96cm}
   \end{center}
   \vspace{-4mm}
   \caption{Nearest-neighbor spacing distribution for the hydrogen atom
            with increasing magnetic field changing from Poisson to Wigner
            function. Taken from Ref.~\cite{hydrogen}.}
   \label{fig5}
\end{figure}
The histogram in Fig.~\ref{fig4} shows the distribution of spacings of nuclear
levels versus the variable $s$, the actual spacing in units of the mean level
spacing $D$. The data set comprises 1726 spacings of levels of the same spin and
parity from a number of different nuclei. These data were obtained from neutron
time-of-flight spectroscopy and from high-resolution proton scattering.
The similarity to Fig.~\ref{fig3} is striking.

\subsection{Hydrogen Atom in a Magnetic Field}

Rydberg levels of the hydrogen atom in a strong magnetic field have a spacing
distribution which once again agrees with RMT, see Fig.~\ref{fig5}~\cite{hydrogen}.
The levels are taken from the vicinity of the scaled binding energy $\tilde E$.
Solid and dashed lines are fits, except for the bottom figure which represents
the GOE. Additionally, a transition from Poisson to GOE behavior with increasing
field strength is clearly visible here.

\subsection{Metal-Insulator Transition}

The conventional way to describe the localization transition is to use the
Anderson Hamiltonian~\cite{Anderson}
\begin{equation}
H = \sum_{i} \epsilon_i a_{i}^{\dagger}a_i \, - \, \sum_{j,i} a_{j}^{\dagger}a_i \ ,
\end{equation}
where $a_{i}^{\dagger}$ and $a_i$ are the electron creation and annihilation
operators at site $i$, subscript $j$ denotes adjacent to $i$, and $\epsilon_i$
is the random energy of site $i$ in units of the overlap energy of adjacent
sites and is uniformly distributed in the range from $[-W/2,W/2]$. 
\begin{figure}[ht]
 \begin{center}
   \hspace{10cm} \psfig{figure=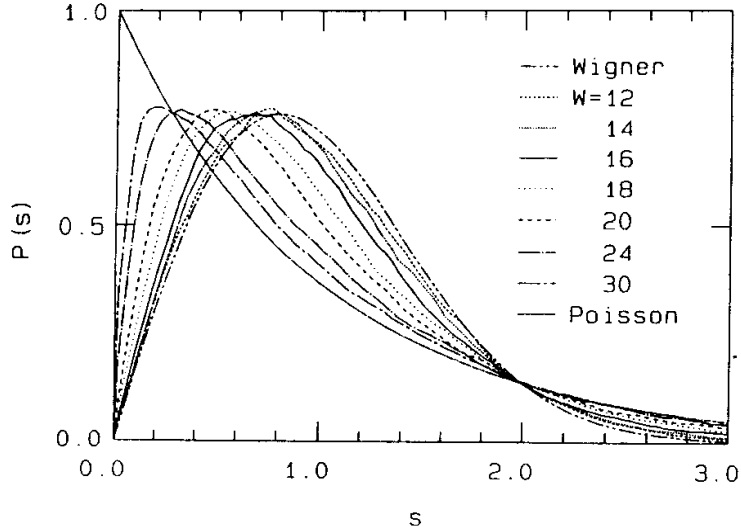,width=12cm,height=8cm}
   \end{center}
   \caption{Nearest-neighbor spacing distribution $P(s)$ for the 3d
            Anderson model at different $W$. Wigner and Poisson functions
            are also shown for comparison. Taken from Ref.~\cite{mesoscopic}.}
   \label{fig6}
\end{figure}
The eigenstates of the Anderson Hamiltonian in the vicinity of the energy 
$\epsilon = 0$ experience a localization transition with increasing $W$.
For a simple cubic lattice, the ``metal-insulator'' transition occurs at
$W = W_c = 16\pm 0.5$ and exhibits a crossover from Wigner to Poisson
behavior, see Fig.~\ref{fig6}. We remark that the Anderson model together
with the Hubbard model serve also as fundamental theories for high-temperature
super-conductivity.

\subsection{Yang-Mills-Higgs System}

The Lagrangian density of the SU(2) Yang-Mills-Higgs (YMH) system is given by
\begin{equation}
{\cal L}^{\rm YMH}={1\over 2}(D_{\mu}\phi )^+(D^{\mu}\phi ) -V(\phi )
-{1\over 4}F_{\mu \nu}^{a}F^{\mu \nu a} \; ,
\end{equation}
where the minimal coupling $D_{\mu}$ of the scalar field $\phi$ to the
gauge field $A_{\mu}^a$ and the field strength tensor $F_{\mu \nu}^{a}$
can be written as
\begin{equation}
(D_{\mu}\phi )=\partial_{\mu}\phi - i g A_{\mu}^b T^b\phi
\; ,
\end{equation}
\begin{equation}
F_{\mu \nu}^{a}=\partial_{\mu}A_{\nu}^{a}-\partial_{\nu}A_{\mu}^{a}+
g\epsilon^{abc}A_{\mu}^{b}A_{\nu}^{c} \; ,
\end{equation}
with $T^b=\sigma^b/2$, $b=1,2,3$, generators of the SU(2) algebra.
The potential of the scalar field (the Higgs field) is
\begin{equation}
V(\phi )=\mu^2 |\phi|^2 + \lambda |\phi|^4 \; .
\end{equation}
This  work is in the (2+1)-dimensional Minkowski space ($\mu =0,1,2$) and
chooses spatially homogeneous Yang-Mills and Higgs fields~\cite{salasnich}.
\begin{figure}[hp]
 \begin{center}
   \hspace{10cm} \psfig{figure=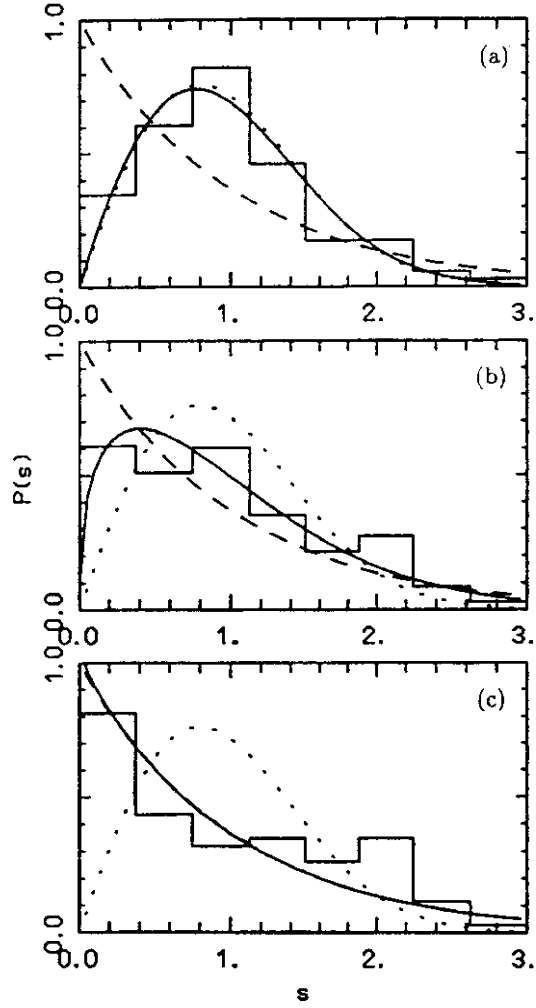,width=9cm,height=14cm}
   \end{center}
   \caption{$P(s)$ distribution for the homogeneous Yang-Mills-Higgs
            system at (a) $v=1$ ($\omega=0.92$), (b) $v=1.1$ ($\omega =0.34$)
            and (c) $v=1.2$ ($\omega =0.01$), where $\omega$ is the Brody parameter
            in Eq.~(\ref{brodydist}).
            First 100 energy levels and interaction $g=1$. The dotted, dashed and
            solid curves stand for Wigner, Poisson and Brody distributions, respectively.
            Taken from Ref.~\cite{salasnich}.}
   \label{fig7}
\end{figure}
When $\mu^2 >0$ the potential $V$ has a minimum at $|{\vec \phi}|=0$,
but for $\mu^2 <0$ the minimum is at
\bear
|{\vec \phi}_0|=\sqrt{-\mu^2\over 4\lambda }=v \ ,
\eear
which is the non-zero Higgs vacuum. This vacuum is degenerate
and after spontaneous symmetry breaking the physical vacuum can be
chosen ${\vec \phi}_0 =(0,0,v)$. If $A_1^1=q_1$, $A_2^2=q_2$
and the other components of the Yang-Mills fields are zero,
in the Higgs vacuum the Hamiltonian of the system reads
\begin{equation}
H={1\over 2}(p_1^2+p_2^2)
+g^2v^2(q_1^2+q_2^2)+{1\over 2}g^2 q_1^2 q_2^2 \; ,
\end{equation}
where $p_1={\dot q_1}$ and $p_2={\dot q_2}$. Here $w^2=2 g^2v^2$ is the
mass term of the Yang-Mills fields. This YMH Hamiltonian is a toy model
for classical non-linear dynamics, exhibiting a classical chaos-order
transition. To outline its connection to the quantal fluctuations of the
energy levels, we plot in Fig.~\ref{fig7} the $P(s)$ distribution for
different values of the parameter $v$. The figure shows a Wigner-Poisson
transition by increasing the value $v$ of the Higgs field in the vacuum.
By using the $P(s)$ distribution and the Brody function it is possible to
give a quantitative measure of the degree of quantal chaoticity of the system.
In this study the influence of the Higgs coupling on the gauge field
is analyzed for a spatially homogeneous YMH system. In the next section we
will address the dependence of the gauge coupling and physical temperature
on the quark field in QCD.

\section{Quantum Chromodynamics}

The Lagrangian ${\cal L}^{\mbox{\scriptsize QCD}}$ of quantum chromodynamics
(QCD) consists of a gluonic part
${\cal L}_{\mbox{\scriptsize G}}^{\mbox{\scriptsize QCD}}$ and a part
${\cal L}_{\mbox{\scriptsize F}}^{\mbox{\scriptsize QCD}}$ from the quarks
\bear
{\cal L}^{\mbox{\scriptsize QCD}} & = &
{\cal L}_{\mbox{\scriptsize G}}^{\mbox{\scriptsize QCD}}+
{\cal L}_{\mbox{\scriptsize F}}^{\mbox{\scriptsize QCD}} \nn \\
& = &
- \frac{1}{4} F_{\mu\nu}^{a}(x)F_{a}^{\mu\nu}(x) +
\sum\limits_{f=1}^{N_{f}} \bar\psi_{f}(x)(i D\!\7 - m_{f}) \psi_{f}(x) \ ,
\eear
with the Dirac spinor $\psi_{f}$, the quark mass $m_{f}$, the number of
flavors $N_{f}$, and the generalized field strength tensor
\bear
F_{a}^{\mu\nu}(x) = \6^{\mu}A_{a}^{\nu}(x)-\6^{\nu}A_{a}^{\mu}(x)
- g f_{abc}A_{b}^{\mu}(x)A_{c}^{\nu}(x)        \ ,
\eear
where the gauge field $A_{a}^{\mu}$ with the SU(3) indices
$a,b,c=1, \dots ,8$, the coupling constant $g$ and the structure
constants $f_{abc}$ of SU(3) enter.
The main object of study is the eigenvalue spectrum of the Dirac
operator of QCD in 4 dimensions
\bear
\FMSlash{D} = \FMSlash{\partial} + i g \FMSlash{A}^a
\frac{\lambda^a}{2} = \gamma_{\mu} \partial_{\mu} + i g \gamma_{\mu} A^a_{\mu}
\frac{\lambda^a}{2} \ ,
\eear
with the $\l_{a}$ the generators of the SU(3) color-group (Gell-Mann matrices).
Discretizing the Dirac operator on a lattice in Euclidean space-time
and applying the Kogut-Susskind (staggered) prescription, leads to the matrix
\bear
(M_{\mbox{\scriptsize KS}})_{xx'}^{aa'} =
\frac{1}{2a} \ \sum_{\mu} \left[
\d_{x+\hat\mu,x'} \ \G_{x\mu} \ U_{x\mu}^{aa'} - \d_{x,x'+\hat\mu}
\ \G_{x'\mu} \ U^{\dagger \ aa'}_{x'\mu} \right] \ ,
\eear
where 
\bear
U_{x \mu} =  \exp \left(igA_{\mu}^{a}(x)\frac{\l^{a}}{2} \ \right) 
\eear
are the gauge field variables on the lattice and $\Gamma_{x\mu}$ a
representation of the $\gamma_{\mu}$-matrices.

In the following we report on work of our own, partly in collaboration
with B.A. Berg, M.-P. Lombardo, and T. Wettig.
In RMT, one has to distinguish several universality classes which are
determined by the symmetries of the system.  For the case of the QCD
Dirac operator, this classification was done in
Ref.~\cite{Verb94}.  Depending on the number of colors and the
representation of the quarks, the Dirac operator is described by one
of the three chiral ensembles of RMT.  As far as the fluctuation
properties in the bulk of the spectrum are concerned, the predictions
of the chiral ensembles are identical to those of the ordinary
ensembles in Sect.~\ref{sec3}~\cite{Fox64}.
In Ref.~\cite{Hala95}, the Dirac matrix was
studied for color-SU(2) using both Kogut-Susskind and Wilson fermions which
correspond to the chiral symplectic (chSE) and orthogonal (chOE) ensemble,
respectively. Here~\cite{Pull98}, we additionally study SU(3) with
Kogut-Susskind fermions which corresponds to the chiral unitary ensemble (chUE).
The RMT result for the nearest-neighbor spacing distribution can be
expressed in terms of so-called prolate spheroidal functions, see
Ref.~\cite{Meht91}.  A very good approximation to $P(s)$ is
provided by the Wigner surmise for the unitary ensemble,
\begin{equation} \label{wigner}
  P_{\rm W}(s)=\frac{32}{\pi^2}s^2e^{-4s^2/\pi} \:.
\end{equation}

We generated gauge field configurations using the standard Wilson
plaquette action for SU(3) with and without dynamical fermions in the
Kogut-Susskind prescription. We have worked on a $6^3\times 4$ lattice
with various values of the inverse gauge coupling $\beta=6/g^2$ both
in the confinement and deconfinement phase.  We typically produced 10
independent equilibrium configurations for each $\beta$.  Because of
the spectral ergodicity property of RMT one can replace ensemble
averages by spectral averages if one is only interested in bulk
properties.

The Dirac operator, $\FMSlash{D}=\FMSlash{\partial}+ig\FMSlash{A}$, is
anti-Hermitian so that the eigenvalues $\lambda_n$ of $i\FMSlash{D}$
are real.  Because of $\{\FMSlash{D},\gamma_5\}=0$ the non-zero
$\lambda_n$ occur in pairs of opposite sign.  All spectra were checked
against the analytical sum rules $\sum_{n} \lambda_n = 0$ and
$\sum_{\lambda_n>0} \lambda_n^2 = 3V$, where V is the lattice volume.
To construct the nearest-neighbor spacing distribution from the
eigenvalues, one first has to ``unfold'' the spectra~\cite{Meht91}.

\begin{figure}[hp]
\begin{center}
\begin{tabular}{ccccc}
  & {\large Confinement $\beta=5.2$}  & \hspace*{6.5mm}   & &
  {\large Deconfinement $\beta=5.4$} \\
  \vspace*{0mm}
  & {\large $ma=0.05$} &&& {\large $ma=0.05$} \\[2mm]
  \multicolumn{2}{c}{\epsfxsize=7cm\epsffile{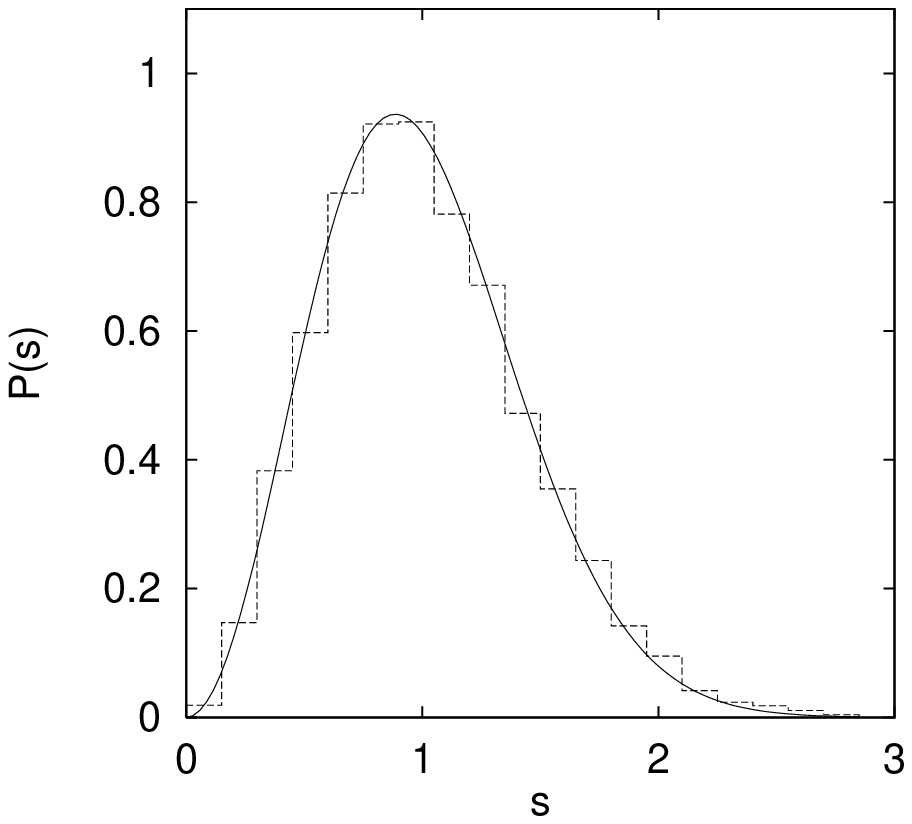}} &&
  \multicolumn{2}{c}{\epsfxsize=7cm\epsffile{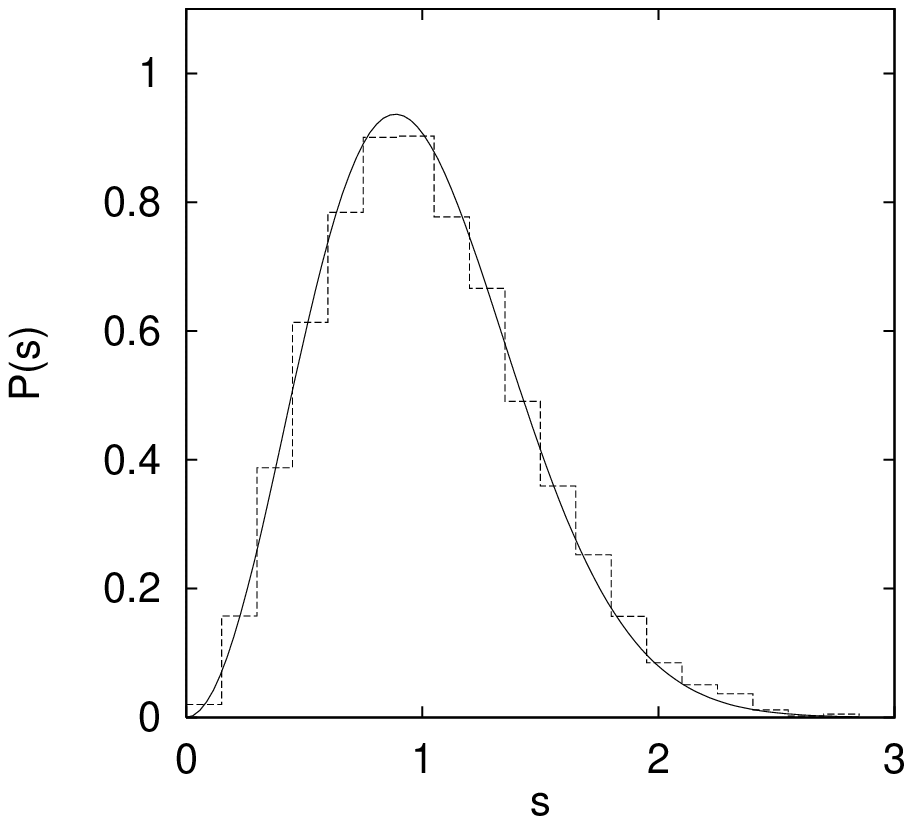}}
\end{tabular}  
\end{center}
\vspace*{-8mm}
\caption{Nearest-neighbor spacing distribution $P(s)$ for the Dirac
  operator on a $6^3 
  \times 4$ lattice in full QCD (histograms) compared with the
  random matrix result (solid lines). There are no changes in $P(s)$
  across the deconfinement phase transition.}
\vspace*{8mm}
\label{fintemp}
  \centerline{
  \psfig{figure=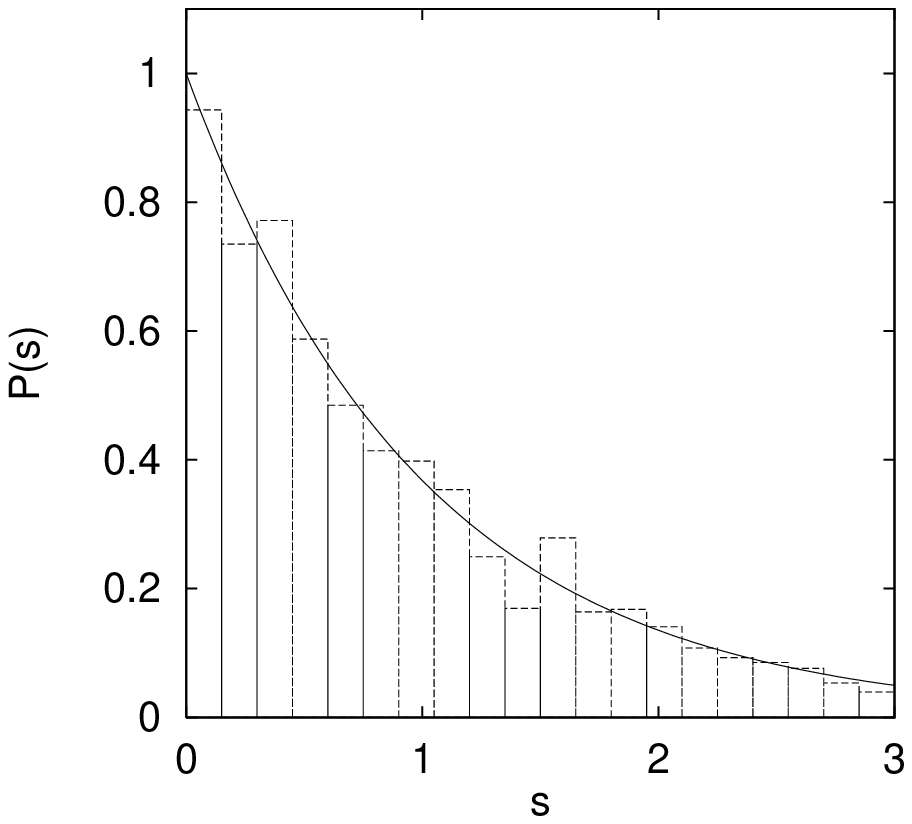,width=7cm}
  }
\vspace*{-5mm}
  \caption{Nearest-neighbor spacing distribution $P(s)$ for the free
            Dirac operator on a $53\times 47\times 43\times 41$ lattice
            compared with a Poisson distribution, $e^{-s}$.}
  \label{free}
\end{figure}

Figure~\ref{fintemp} compares $P(s)$ of full QCD with $N_f = 3$
flavors and quark mass $ma=0.05$ to the RMT result.  In the confinement
as well as in the deconfinement phase we observe agreement with RMT up
to very high $\beta$ (not shown).  The observation that $P(s)$ is not
influenced by the presence of dynamical quarks is
expected from the results of Ref.~\cite{Fox64}, which
apply to the case of massless quarks. Our
results, and those of Ref.~\cite{Hala95}, indicate that massive
dynamical quarks do not affect $P(s)$ either.

No signs for a transition to Poisson regularity are found. The
deconfinement phase transition does not seem to coincide with a
transition in the spacing distribution. For very large values of
$\beta$ far into the deconfinement region, the eigenvalues
start to approach the degenerate eigenvalues of the free theory, given
by $\lambda^2=\sum_{\mu=1}^4 \sin^2(2\pi n_\mu/L_\mu)/a^2$, where $a$
is the lattice constant, $L_{\mu}$ is the number of lattice sites in
the $\mu$-direction, and $n_\mu=0,\ldots,L_\mu-1$.  In this case, the
nearest-neighbor spacing distribution is neither Wigner nor Poisson.
It is possible to lift the degeneracies of the free
eigenvalues using an asymmetric lattice where $L_x$, $L_y$, etc. are
relative primes and, for large lattices, the distribution 
is then Poisson, $P_{\rm P}(s)=e^{-s}$, see Fig.~\ref{free}.

\begin{figure*}[t]
  \centerline{\psfig{figure=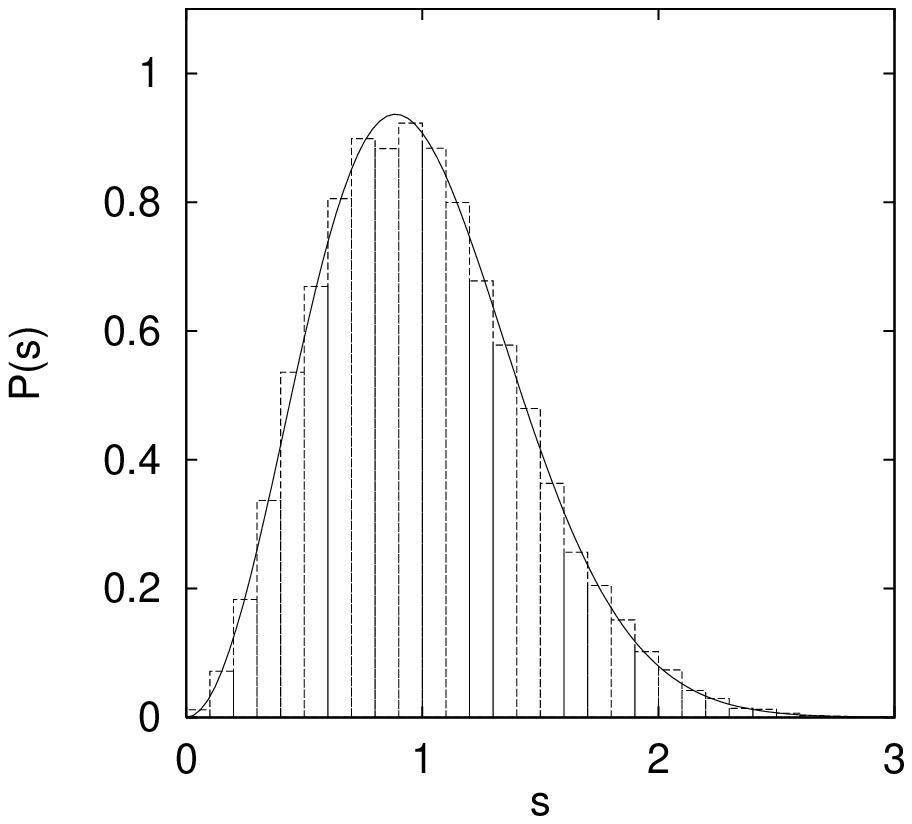,width=7cm}\hspace*{10mm}
    \psfig{figure=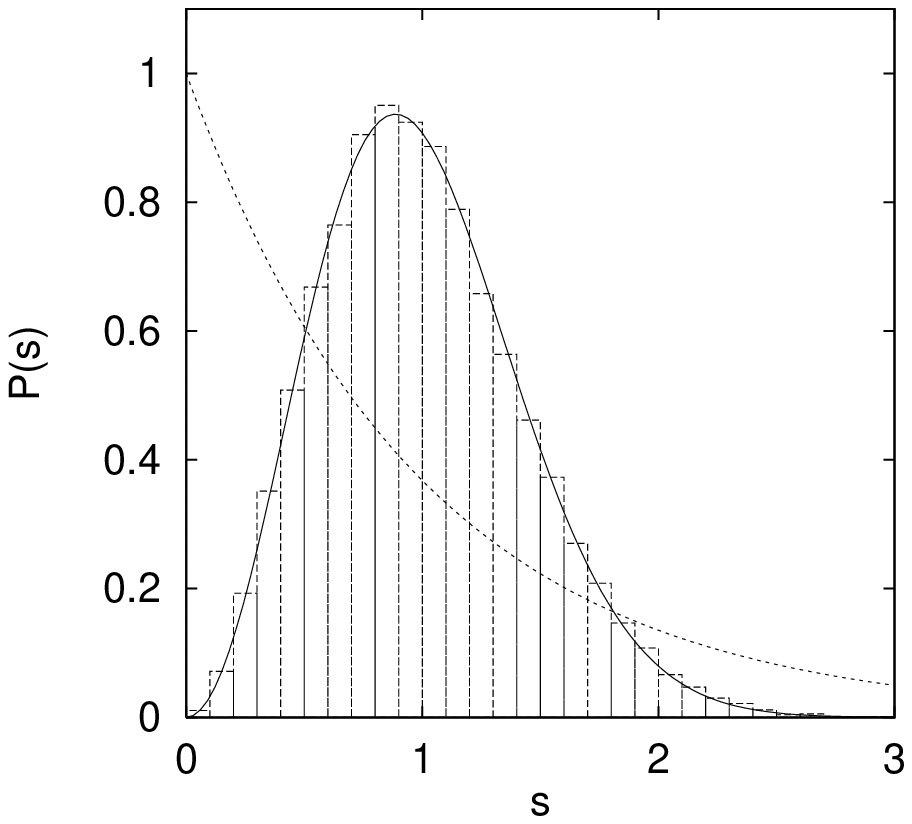,width=7cm}}
  \caption{Nearest-neighbor spacing distribution $P(s)$ for U(1) gauge
    theory on an $8^3\times 6$ lattice in the confined phase (left)
    and in the Coulomb phase (right). The theoretical curves are the chUE
    result, Eq.~(\ref{wigner}), and the Poisson distribution, $P_{\rm
      P}(s)=\exp(-s)$.}
  \label{f02}
\end{figure*}

We have also investigated the staggered Dirac spectrum of 4d U(1)
gauge theory which corresponds to the chUE of RMT but had
not been studied before in this context.  At $\beta_c \approx 1.01$
U(1) gauge theory undergoes a phase transition between a confinement
phase with mass gap and monopole excitations for $\beta < \beta_c$ and
the Coulomb phase which exhibits a massless photon for $\beta >
\beta_c$~\cite{BePa84}. As for SU(2) and SU(3) gauge groups, we
expect the confined phase to be described by RMT, whereas free
fermions are known to yield the Poisson distribution (see
Fig.~\ref{free}). The question arose whether the Coulomb phase would be
described by RMT or by the Poisson distribution~\cite{BeMaPu99}. The
nearest-neighbor spacing distributions for an $8^3\times 6$ lattice at
$\beta=0.9$ (confined phase) and at $\beta=1.1$ (Coulomb phase),
averaged over 20 independent configurations, are depicted in
Fig.~\ref{f02}. Both are consistent with the chUE of RMT.

Physical systems which are described by non-Hermitian operators have
attracted a lot of attention recently, among others QCD at non-zero
chemical potential $\mu$~\cite{Step96}.
A formulation of the QCD Dirac
operator at $\mu\ne0$ on the lattice in the staggered scheme is
given by \cite{Hase83}
\begin{equation}
\label{Dirac}
M_{x,y}(U,\mu) = \frac{1}{2a} \left\{
\sum\limits_{\nu=\hat{x},\hat{y},\hat{z}}
  \left[U_{x \nu} \Gamma_{x \nu} \delta_{y,x\!+\!\nu}-{\rm h.c.}\right]
+ \left[U_{x \hat{t}} \Gamma_{x \hat{t}} e^{\mu}
    \delta_{y,x\!+\!\hat{t}}
    -U_{y \hat{t}}^{\dagger} \Gamma_{y \hat{t}}
    e^{-\mu}\delta_{y,x\!-\!\hat{t}}\right] \right\} \ ,
\end{equation}
with the link variables $U$ and the staggered phases $\Gamma$.
For $\mu>0$, the Dirac operator
loses its Hermiticity properties so that its eigenvalues become
complex. The aim of the present analysis is to investigate whether
non-Hermitian RMT is able to describe the fluctuation properties of
the complex eigenvalues of the QCD Dirac operator.  The eigenvalues
are generated on the lattice for various values of $\mu$.  We apply a
two-dimensional unfolding procedure~\cite{Mark99} to separate the
average eigenvalue 
density from the fluctuations and construct the nearest-neighbor
spacing distribution, $P(s)$, of adjacent eigenvalues in the complex
plane. Adjacent eigenvalues are defined to be the pairs for which the
Euclidean distance in the complex plane is smallest.  
The data are then compared to analytical predictions of the Ginibre
ensemble \cite{Gini65} of non-Hermitian RMT, which describes the
situation where the real and imaginary parts of the strongly
correlated eigenvalues have approximately the same average magnitude.
In the Ginibre ensemble, the average spectral density is already
constant inside a circle and zero outside.  In this
case, unfolding is not necessary, and $P(s)$ is given by \cite{Grob88}
\begin{equation} \label{Ginibre}
  P_{\rm G}(s)  =  c\, p(cs)\:, ~~p(s) = 
  2s\lim_{N\to\infty}\left[\prod_{n=1}^{N-1}e_n(s^2)\,e^{-s^2}
  \right] \sum_{n=1}^{N-1}\frac{s^{2n}}{n!e_n(s^2)}\:,
\end{equation}
where $e_n(x)=\sum_{m=0}^n x^m/m!$ and $c=\int_0^\infty ds \, s \,
p(s)=1.1429...$. 
For uncorrelated eigenvalues in the
complex plane, the Poisson distribution becomes \cite{Grob88}
\begin{equation}
  \label{Poisson}
  P_{\bar{\rm P}}(s)=\frac{\pi}{2}\,s\,e^{-\pi s^2/4}\:.
\end{equation}
This should not be confused with the Wigner distribution~(\ref{wigner}).

We report on simulations done with gauge group SU(2) on a $6^4$ lattice
using $\beta=4/g^2=1.3$ in the confinement region for $N_f=2$ flavors
of staggered fermions of mass $ma=0.07$. For this system the fermion
determinant is real and lattice simulation become feasible~\cite{Hand99}.
We sampled 160 independent configurations~\cite{lattice}.
In the case of color-SU(2), the
staggered Dirac operator has an extra anti-unitary symmetry
\cite{Hands90} and falls in the symmetry class with Dyson parameter
$\beta_D=4$~\cite{Hala97b}. However, one can show that the
nearest-neighbor spacing distribution in the bulk of the spectrum
for this class is also given by Eq.~(\ref{Ginibre}).

\begin{figure}
\begin{center}
\begin{tabular}{ccccc}
  & {\large $\mu=0$}  & \hspace*{6.5mm}   & &
  {\large $\mu=0.4$} \\
  \vspace*{0mm}
  & $ma=0.07$ &&& $ma=0.07$ \\[2mm]
  \multicolumn{2}{c}{\epsfxsize=7cm\epsffile{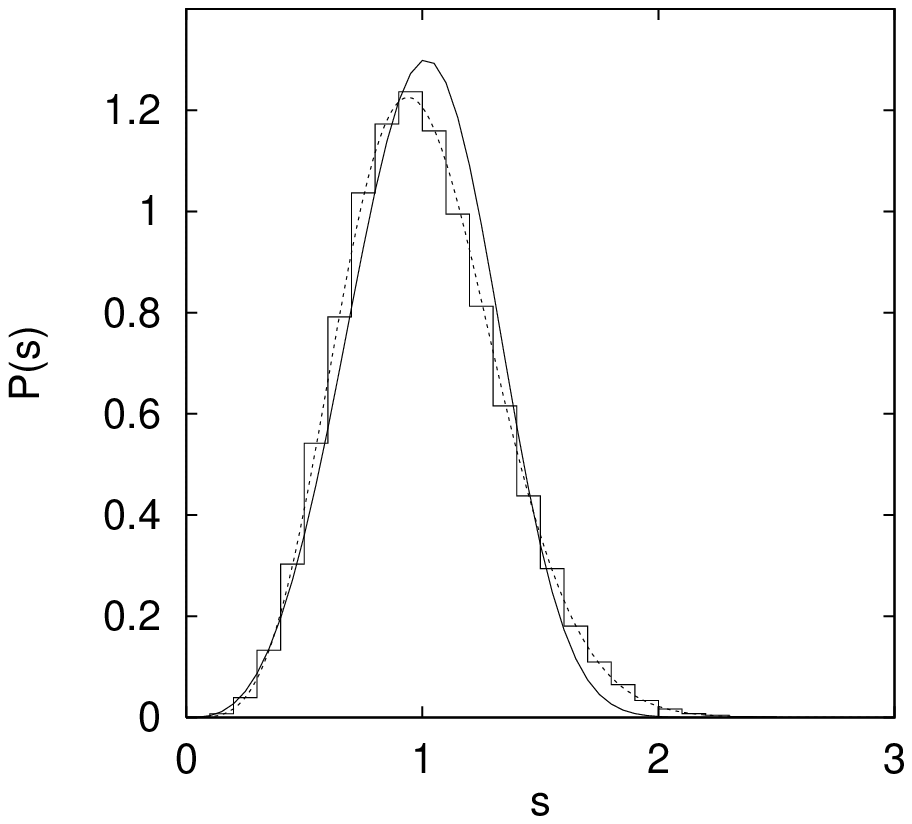}} &&
  \multicolumn{2}{c}{\epsfxsize=7cm\epsffile{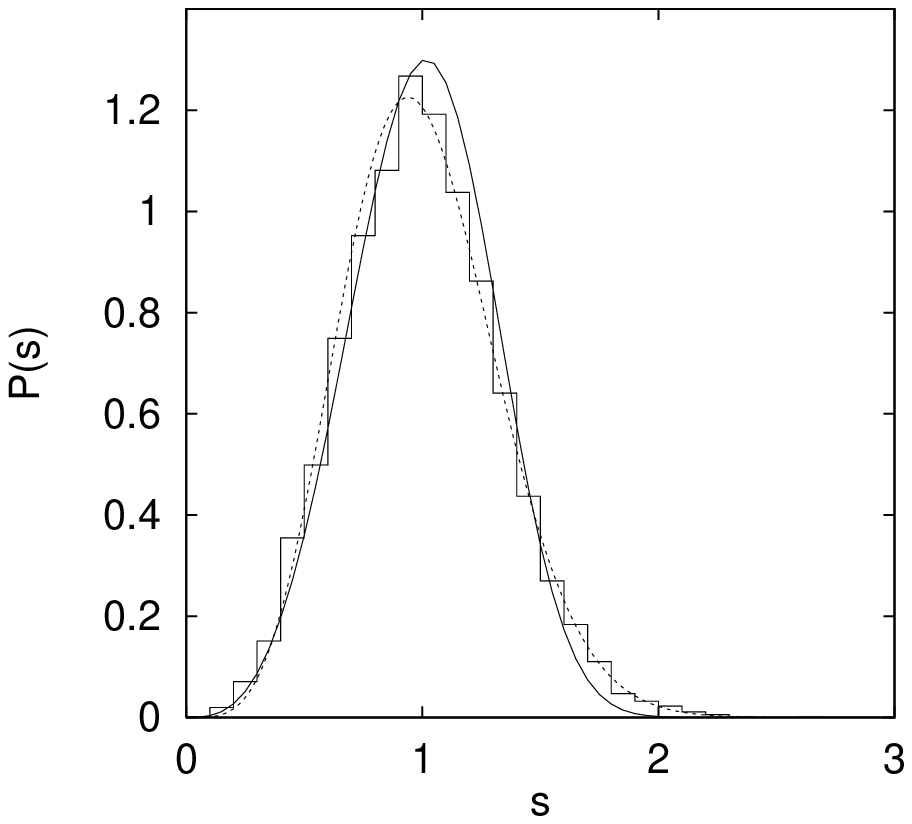}}
\end{tabular}
\end{center}
\vspace*{-3mm}
  \caption{Nearest-neighbor spacing distribution in the complex plane
    for two-color QCD with 
    $\mu$ in the confinement (left) and deconfinement (right) phase.
    The short-dashed curve is the Wigner distribution
    for the chSE and the solid curve is the Ginibre distribution of
    Eq.~(\protect\ref{Ginibre}).}
  \vspace*{-3mm}
  \label{Psconf}
\end{figure}

Our results for $P(s)$ are presented in Fig.~\ref{Psconf}. As a function
of the chemical potential $\mu$, we expect to find a transition from Wigner
to Ginibre behavior in $P(s)$. This was clearly seen in color-SU(3) with
$N_f=3$ flavors and quenched chemical potential~\cite{Mark99}, where
differences between both curves are more pronounced. For the
symplectic ensemble of color-SU(2) with staggered fermions, the Wigner
and Ginibre distributions are very close to each other and thus harder
to distinguish. They are reproduced by our preliminary data for
$\mu=0$ and $\mu=0.4$, respectively.

\section{Conclusion}

We have outlined the universal applicability of random-matrix theory.
Several examples of quantum chaos from the literature have been shown
~\cite{Guhr}, both from numerical simulation and physical experiment.
In some of these examples the transition from regularity to chaoticity
could be observed, both for the classical and quantum system.

Concerning our own studies of quantum chromodynamics, we were able to
demonstrate that the nearest-neighbor spacing distribution $P(s)$ of the
eigenvalues of the Dirac operator agrees perfectly with the RMT prediction
both in the confinement and quark-gluon plasma-phase. This means that
QCD is governed by quantum chaos in both phases. We could show that the
eigenvalues of the free Dirac theory yield a Poisson distribution related
to regular behavior. Our investigations tell us that the critical point of
the spontaneous breaking of chiral symmetry does not coincide with a
chaos-to-order transition.

\section{Acknowledgments}
This study was supported in part by FWF project P14435-TPH.
We thank B.A. Berg, M.-P. Lombardo, and T. Wettig for collaborations.


\begin{thebibliography}{99}
\bibitem{schuster} H.G. Schuster, {\it Deterministic Chaos: an Introduction}
  (VCH, Weinheim, 1995).
\bibitem{Guhr} T. Guhr, A. M\"uller-Groeling, and H.A. Weidenm\"uller,
  Phys. Rep. 299 (1998) 189.
\bibitem{McDoKauf} S.W. McDonald and A.N. Kaufman, Phys. Rev. Lett. 42
  (1979) 1189.
\bibitem{Casa} G. Casati, F. Valz-Gris, and I. Guarneri, Lett. Nuovo Cimento
  28 (1980) 279.
\bibitem{Berr} M.V. Berry, Ann. Phys. (NY) 131 (1981) 163.
\bibitem{Robn} M. Robnik, J. Phys. A 17 (1984) 1049.
\bibitem{SeVeZi} T.H. Seligman, J.J.M. Verbaarschot, and M.R. Zirnbauer,
    Phys. Rev. Lett. 53 (1984) 215;
    T.H. Seligman, J.J.M. Verbaarschot, and M.R. Zirnbauer,
    J. Phys. A 18 (1985) 2751.
\bibitem{Bohi84} O. Bohigas, M.-J. Giannoni, and C. Schmit,
  Phys. Rev. Lett. 52 (1984) 1.
\bibitem{Brod} T.A. Brody, Lett. Nuovo Cimento 7 (1973) 482.
\bibitem{nuclear} O. Bohigas, R.U. Haq, and A. Pandey, in {\it Nuclear
  Data for Science and Technology}, K.H. B\"ochhoff (Ed.) (Reidel, Dordrecht,
  1983) p. 809.
\bibitem{hydrogen} D. Wintgen and H. Friedrich, Phys. Rev. A 35 (1987) 1464.
\bibitem{Anderson} P.W. Anderson, Phys. Rev. 109 (1958) 1492.
\bibitem{mesoscopic}  B.L. Al'tshuler, I.Kh. Zharekeshev, S.A.
  Kotochigova, and B.I. Shklovski$\breve\i$,
  Zh. Eksp. Teor. Fiz. 94 (1988) 343 [Sov. Phys. JETP 67 (1988) 625];
  B.I. Shklovski$\breve\i$, B. Shapiro, B.R. Sears, P. Lambrianides,
  and H.B. Shore, Phys. Rev. B 47 (1993) 11487.
\bibitem{salasnich} L.\ Salasnich, Mod.\ Phys.\ Lett.\ A 12 (1997) 1473.
\bibitem{Verb94} J.J.M.\ Verbaarschot, Phys.\ Rev.\ Lett.\ 72 (1994)
  2531.
\bibitem{Fox64} D. Fox and P.B. Kahn, Phys. Rev. 134 (1964) B1151;
  T. Nagao and M. Wadati, J. Phys. Soc. Jpn. 60 (1991) 3298;
  61 (1992) 78;
  61 (1992) 1910.
\bibitem{Hala95} M.A.\ Halasz and J.J.M.\ Verbaarschot, Phys.\ Rev.\
  Lett.\ 74 (1995) 3920;
  M.A.\ Halasz, T.\ Kalkreuter, and J.J.M.\
  Verbaarschot, Nucl.\ Phys.\ B (Proc.\ Suppl.) 53 (1997) 266.
\bibitem{Pull98} R. Pullirsch, K. Rabitsch, T. Wettig, and H. Markum,
  Phys. Lett. B 427 (1998) 119.
\bibitem{Meht91} M.L.\ Mehta, {\it Random Matrices}, 2nd Ed. (Academic
  Press, San Diego, 1991).
\bibitem{BePa84} B.A. Berg and C. Panagiotakopoulos, Phys. Rev. Lett.
  52 (1984) 94.
\bibitem{BeMaPu99} B.A. Berg, H. Markum, and R. Pullirsch,
  Phys. Rev. D 59 (1999) 097504.
\bibitem{Step96} M.A. Stephanov, Phys. Rev. Lett. 76 (1996) 4472.
\bibitem{Hase83} P. Hasenfratz and F. Karsch, Phys. Lett. B 125
  (1983) 308;
  J. Kogut, H. Matsuoka, M. Stone, H.W. Wyld, S. Shenker,
  J. Shigemitsu, and D.K. Sinclair, Nucl. Phys. B 225 (1983) 93;
  I.M. Barbour, Nucl. Phys. B (Proc. Suppl.) 26 (1992) 22.
\bibitem{Mark99} H. Markum, R. Pullirsch, and T. Wettig, Phys. Rev.
  Lett. 83 (1999) 484.
\bibitem{Gini65} J. Ginibre, J. Math. Phys. 6 (1965) 440.
\bibitem{Grob88} R. Grobe, F. Haake, and H.-J. Sommers, Phys. Rev.
  Lett. 61 (1988) 1899.
\bibitem{Hand99} S. Hands, J.B. Kogut, M.-P. Lombardo, and S.E. Morrison,
  Nucl. Phys. B 558 (1999) 327.
\bibitem{lattice} E. Bittner, M.-P. Lombardo, H. Markum, and R. Pullirsch,
 Nucl. Phys. B (Proc. Suppl.) 94 (2001) 445.
\bibitem{Hands90} S. Hands and M. Teper, Nucl. Phys. B 347 (1990) 819.
\bibitem{Hala97b} M.A. Halasz, J.C. Osborn, and J.J.M. Verbaarschot,
  Phys. Rev. D 56 (1997) 7059.
\end{thebibliography}
\end{document}